\newcommand{\be}{\begin{equation}}
\newcommand{\ee}{\end{equation}}

\def\Ok{\Omega_{\rm k}}
\def\Ode{\Omega_{\rm de}}

\newcommand{\Omegak}{$\Omega_{\rm k}$}

\newcommand\rxj{RXJ1131$-$1231}

\documentclass{iau-JDSS-hacked}
\usepackage{natbib}
\usepackage{epsfig}
\usepackage{psfig}
\usepackage{graphicx}
\usepackage{sidecap}
\usepackage{amsmath}
\usepackage[usenames]{color}

\usepackage{natbib}
\def\reff@jnl#1{{\rm#1\/}}
\def\aj{\reff@jnl{AJ}}                  % Astronomical Journal
\def\araa{\reff@jnl{ARA\&A}}            % Annual Review of Astron and Astrophys
\def\apj{\reff@jnl{ApJ}}                % Astrophysical Journal
\def\apjl{\reff@jnl{ApJ}}               % Astrophysical Journal, Letters
\def\apjs{\reff@jnl{ApJS}}              % Astrophysical Journal, Supplement
\def\apss{\reff@jnl{Ap\&SS}}            % Astrophysics and Space Science
\def\aap{\reff@jnl{A\&A}}               % Astronomy and Astrophysics
\def\aapr{\reff@jnl{A\&A~Rev.}}         % Astronomy and Astrophysics Reviews
\def\aaps{\reff@jnl{A\&AS}}             % Astronomy and Astrophysics, Supplement
\def\mnras{\reff@jnl{MNRAS}}            % Monthly Notices of the RAS
\def\prd{\reff@jnl{Phys.Rev.D}}         % Physical Review D
\def\prl{\reff@jnl{Phys.Rev.Lett}}      % Physical Review Letters
\def\pasp{\reff@jnl{PASP}}              % Publications of the ASP
\def\pasj{\reff@jnl{PASJ}}              % Publications of the ASJ
\def\nat{\reff@jnl{Nature}}             % Nature 

\title[] %% give here short title %%
{Dark energy with gravitational lens time delays}

\author[] %% give here short author list %%
{ T.~Treu$^1$, P.J.~Marshall$^{2}$, F.-Y., Cyr-Racine$^3$,
C.D.~Fassnacht$^4$, C.R.~Keeton$^5$,
E.V.~Linder$^6$,L.A.~Moustakas$^3$, M.~Brada\v{c}$^4$,
E.~Buckley-Geer$^7$, T.~Collett$^8$, F.~Courbin$^9$, G.~Dobler$^1$,
D.A.~Finley$^7$, J.~Hjorth$^{10}$, C.S.Kochanek$^{11}$,
E.~Komatsu$^{12}$, L.V.E.~Koopmans$^{13}$, G.~Meylan$^9$,
P.~Natarajan$^{14}$, M.~Oguri$^{15}$, S.H.~Suyu$^{16}$, M.~Tewes$^9$,
K.C.~Wong$^{17}$, A.I.~Zabludoff$^{17}$, D.~Zaritsky$^{17}$,
T.~Anguita$^{18}$, R.J.~Brunner$^{19}$, R.~Cabanac$^{20}$,
E.E.~Falco$^{21}$, A.~Fritz$^{22}$, G.~Seidel$^{23}$,
D.A.~Howell$^{24}$, C.~Giocoli$^{25}$, N.~Jackson$^{26}$,
S.~Lopez$^{27}$, R.B.~Metcalf$^{25}$, V.~Motta$^{28}$,
T.~Verdugo$^{29}$}

\affiliation{$^1$University of California Santa Barbara, email:
  tt@physics.ucsb.edu; 
$^2$Kavli Institute for Particle Astrophysics and Cosmology, Stanford University;
$^3$ Jet Propulsion Laboratory;
$^4$ University of California Davis;
$^5$ Rutgers University;
$^6$Lawrence Berkeley National Laboratory; 
$^7$ Fermilab Center for Particle Astrophysics;
$^8$ Cambridge University, UK;
$^9$ Ecole Polytechnique F\'ed\'erale de Lausanne. CH;
$^{10}$ Dark Cosmology Centre, University of Copenhagen, D;
$^{11}$ Ohio State University;
$^{12}$ Max-Planck-Institut f\"ur Astrophysik, Garching, Germany;
$^{13}$ Kapteyn Institute, Groningen, the Netherlands; 
$^{14}$ Yale University;
$^{15}$ Kavli IPMU (WPI), University of Tokyo, Japan;
$^{16}$ ASIAA, Taiwan;
$^{17}$ University of Arizona;
$^{18}$ Universidad Andres Bello, Chile;
$^{19}$ University of Illinois;
$^{20}$ Universit\`e de Toulouse, France;
$^{21}$ Smithsonian Astrophysical Observatory;
$^{22}$ INAF Milano, Italy;
$^{23}$ Max Planck Institute f\"ur Astronomie, Heidelberg, Germany;
$^{24}$ Las Cumbres Observatory Global Telescope Network;
$^{25}$ Universit\`a di Bologna, Italy;
$^{26}$ Jodrell Bank, UK;
$^{27}$ Universidad de Chile, Chile;
$^{28}$ Universidad de Valparaiso, Chile;
$^{29}$ Centro de Investigaciones de Astronom\'ia, Venezuela;}

\pubyear{2013}
\volume{Volume 1}  %% insert here IAU Highlights of Astronomy Volume Number
\pagerange{1-2}
\date{}
\jname{SNOWMASS-2013}
\editors{}
\begin{document}

\maketitle

\begin{abstract} Strong lensing gravitational time delays are a powerful and
cost effective probe of dark energy. Recent studies have shown that a
single lens can provide a distance measurement with 6-7\% accuracy
(including random and systematic uncertainties), provided sufficient
data are available to determine the time delay and reconstruct the
gravitational potential of the deflector. Gravitational-time delays
are a low redshift ($z\sim 0-2$) probe and thus allow one to break
degeneracies in the interpretation of data from higher-redshift probes
like the cosmic microwave background in terms of the dark energy equation
of state. Current studies are limited by the size of the sample of
known lensed quasars, but this situation is about to change. Even in
this decade, wide field imaging surveys are likely to discover
thousands of lensed quasars, enabling the targeted study of $\sim100$ of these
systems and resulting in substantial gains in the dark energy figure
of merit.  In the next decade, a further order of magnitude
improvement will be possible with the $10^4$ systems expected to be
detected and measured with LSST and Euclid. To fully exploit these
gains, we identify three priorities. First, support for the
development of software required for the analysis of the data.
Second, in this decade, small robotic telescopes (1-4m in diameter)
dedicated to monitoring of lensed quasars will transform the field by
delivering accurate time delays for $\sim100$ systems. Third, in the
2020's, LSST will deliver 1000's of time delays; the bottleneck will
instead be the aquisition and analysis of high resolution imaging
follow-up. Thus, the top priority for the next decade is to support
fast high resolution imaging capabilities, such as those enabled by
the James Webb Space Telescope and next generation adaptive optics
systems on large ground based telescopes.

%\keywords{cosmology: distance scale - cosmology: cosmological parameters - cosmology: dark energy, galaxies: distances and redshifts}

\end{abstract}

\section{Executive Summary}

Strong gravitational lensing time delays measure distances, and hence
the Hubble constant, dark energy, and dark matter.  They measure
cosmographic distance ratios that are highly complementary to those
measured by supernovae and baryon acoustic oscillations and break many
of the degeneracies inherent in the interpretation of cosmic microwave
background data. They require no new major facilities, simply support
for development and execution of sophisticated data analysis
techniques to detect, measure and model the lenses, re-purposing and
robotization of existing telescopes, and support of adaptive
optics technology.

The priorities to support strong lensing time delays as a dark energy
probe are:

\begin{enumerate}
\item Support the development of software and methods to ensure maximal information gain from the planned wide field, cadenced 
survey imaging data, and accurate inferences from them;
\item Transformation of a 2-4m class telescope (or a network of 1m
telescopes) into a high cadence, long term dark energy monitoring
experiment (non-exclusive use);
\item Support development of high performance
adaptive optics systems on 8-30m class telescopes.
\end{enumerate}

This program leverages and enhances existing surveys such as
Pan-STARRS1, Dark Energy Survey, and Hyper Suprime-Cam which will
discover $>100$ well-measurable time delay systems this decade, and
LSST which will discover $>1000$ systems next decade, and involves
high performance computing delivering added value to the analysis
pipelines of these experiments.

\section{Introduction}

\begin{figure*}
  \renewcommand{\baselinestretch}{0.7} \centering
  \includegraphics[width=0.7\textwidth]{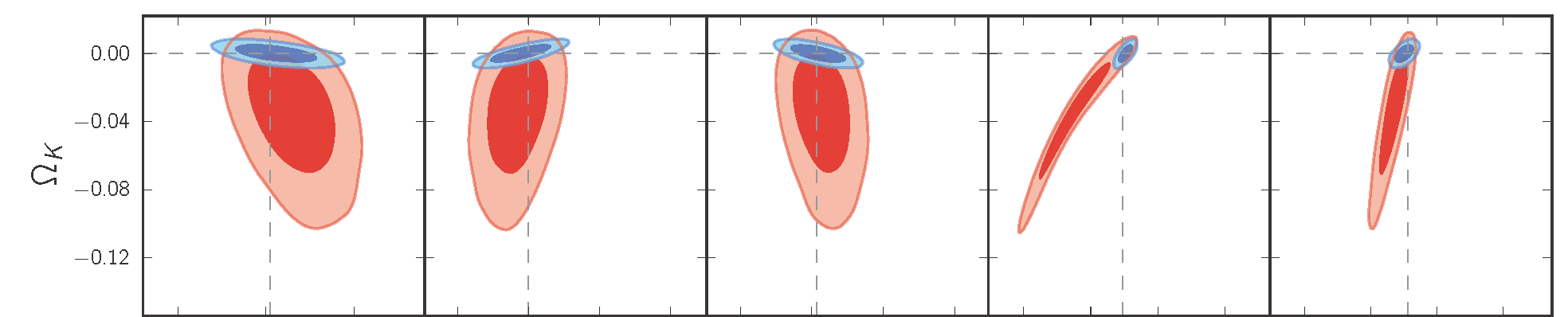} \centering
  \includegraphics[width=0.7\textwidth]{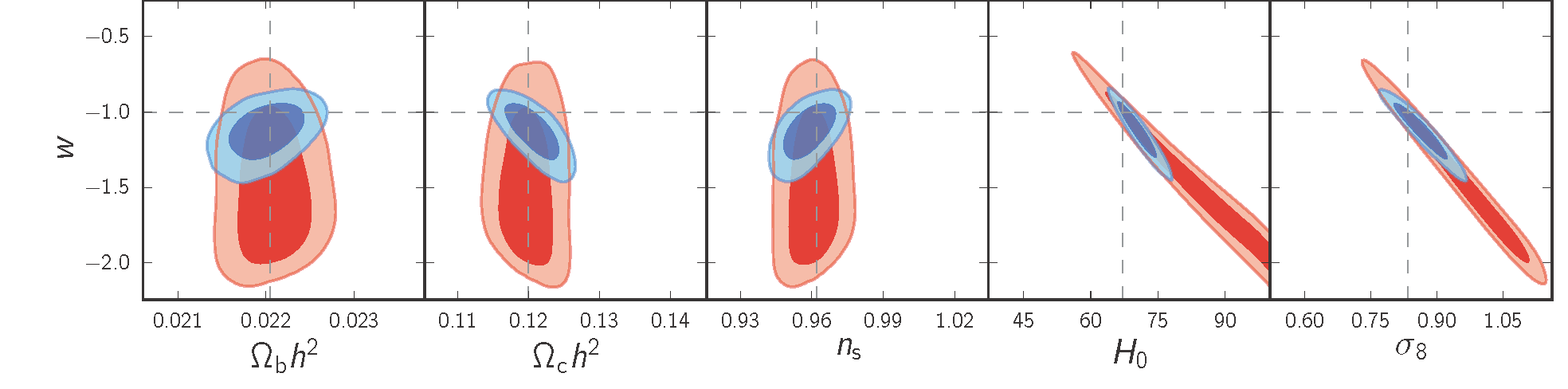}
  \caption{\label{fig:Planck16} Cosmological inference based on the
  Planck data alone (red contours) and in combination with a low
  redshift probe (Baryonic Acoustic Oscillatios; in blue). Note how
  for generic models CMB data alone cannot constrain simultaneously
  \Omegak, $w$, and $H_0$, because the main observable
  provided by CMB data is the angular diameter distance to the epoch
  of recombination. Only when combining with a low redshift probe,
  like BAO, time delay lenses, or local measurements of the Hubble
  constant, the degeneracies can be broken and one can determine $w$
  and curvature. From Planck Paper XVI \citep{Planck16}.}
\end{figure*}

The acceleration of the universe is one of the most profound mysteries
in physics \citep{Rie++98,Per++99}. Understanding its origin may lead
to a rethinking of the standard paradigm of the cosmological model
based on general relativity and cold dark matter.  Ongoing and
upcoming studies of dark energy rely heavily on the accurate knowledge
of distances in the nearby universe (redshift $z\lesssim1$). This is
illustrated very clearly by the recent Planck results shown in
Figure~\ref{fig:Planck16}. The anisotropies of the cosmic microwave
background are primarily sensitive to the angular diameter distance to
the epoch of recombination ($z\sim1000$) and therefore contain little
information of later time phenomena like dark energy. This results in
substantial degeneracies between the equation of state parameter
($w$), curvature (\Omegak), and the Hubble constant ($H_0$). The
addition of CMB lensing information mitigates but does not solve the
problem \citep{Planck16}. Only with the addition of lower redshift
probes, like gravitational time delays (or Baryonic Acoustic
Oscillations or local distance ladder measurements of $H_0$) these
degeneracies can be broken allowing one to learn about the nature of
dark energy.

In order to reach the goal of the next decade's cosmological
experiments, it will be necessary to pin down the accuracy on local
distances, and thus equivalently $H_0$, below the percent level
\citep{Riess++11,Fre++12}. Moreover, to identify unknown systematic
errors in existing techniques, it is essential to gather several
independent measurements \citep{Wei++12, Lin11}. The tension between
previous measurements of $H_0$ and that recently derived by the Planck
team within the assumptions of a six-parameter flat $\Lambda$CDM model
(including tension with WMAP9) highlights the need for multiple
independent measurements. If the tension cannot be resolved by unknown
systematics, it will force the rejection of the six-parameter model in
favor of a more complex alternative, thus leading to new physics such
as a non-trivial dark energy equation of state or alternative theories
of gravity. Even if the current tension can be resolved by discovering
unknown systematics, adding a lower redshift measurement is essential
to probe dark energy when it becomes relevant.
 
The gravitational time delay technique \citep{Ref64}, applied to a
large number of lensed systems, is one of only a few that can lead to
subpercent accuracy on low redshift distances and therefore $H_0$. By
measuring the time delay $\Delta t$ between pairs of strongly
gravitationally lensed images, and modeling the mass distribution of
the lens galaxy, the time delay distance $D_{\Delta t}$ can be
inferred \footnote{The time delay distance $D_{\Delta t} \equiv (1 +
z_d) D_d D_s /D_{ds}$ is a ratio of angular diameter distances ($D$;
$s$=source, $d$=deflector) and contains all the cosmological
information \citep[see, e.g.,][for a description]{Tre10}}. Being a physical
distance (as opposed to a relative distance modulus), this quantity is
primarily sensitive to $H_0$. However, samples of lenses also contain
cosmological information in the form of distance ratios
\citep[e.g.,][and references therein]{Ogu07,C+M09}. In the past, the
technique was plagued by poor time delay measurements, invalid
assumptions about the lens mass profile, and systematic errors
associated with over-simplistic modelling of the mass distribution in
and around the lens.  However, times have changed.  It has been
recently demonstrated that a single gravitational lens with
well-measured time delays can be used to measure time delay distances
to 6-7\% total uncertainty \citep[random {\it and}
systematic][]{Suy++10,Suy++13}. The key breakthroughs required to
achieve this were: 1) multi-year monitoring to determine time delays
\citep[][Figure 2]{Fas++02,Koc++06a,Tew++12}, 2) the use of high resolution imaging and
stellar kinematics to pinpoint the gravitational potential of the lens
using advanced modeling techniques
\citep[][Figure 3]{T+K02b,Suy++10,Suy++13}, and 3) accounting in detail for the mass
distribution along the line of sight \citep{Suy++10,Fas++11,Gre++13,Col++13}.
The analysis of just two systems produced measurements of $H_0$, $w$, and
\Omegak\ that are competitive with and highly complementary
to those from established methods such as Cepheids, Baryonic Acoustic
Oscillations and Supernovae Ia
(Fig.~\ref{fig:Lenses_BAO_SN}). Gravitational lens time delays are not
only independent of these other methods, but are also a cost-effective
probe of the dark energy equation of state, as described below.

\section{How Does Time Delay Lens Cosmography Work?}  

According to Fermat's principle, images form at the extrema of the
arrival time surface of any lens -- including gravitational ones.  The
time delay between multiple images can be measured by monitoring a
variable source like a quasar. In short, the differerence in arrival
time is given by $\Delta t \propto D_{\Delta t}{\Delta \phi}$. Here,
the time delay distance contains all of the cosmological dependence,
while the Fermat potential $\phi$ depends only on the details of the
mass distribution.  By measuring a time delay and determining a mass
model for the main deflector, one obtains the time delay distance
$D_{\Delta t}$ and, thus, a determination of cosmological parameters
\citep{Ref64,Sch++97,T+K02b,Koc02,Koo++03,Ogu07,Vui++08,Suy++10,P+H10,Suy++13}.
Time delays constrain cosmology in two ways. First, they pin down the
Hubble constant and thus remove degeneracies in the interpretation of
CMB data in terms of $w$ and its evolution
\citep[][Figure~5]{Lin11}. Second, a sample of lenses at different
redshifts measures angular distance ratios and thus is directly
sensitive to other parameters
\citep{C+M09}, especially curvature~\citep[][see Figure~4]{Suy++13}.  The strengths of time delay lens cosmography are
that (1) the method is based on simple geometry and well-tested
physics (i.e., general relativity) rather than complicated
astrophysics (e.g., supernova explosions, structure formation, etc.)
that may not be completely understood, and (2) it produces a direct
physical measurement of a cosmological distance.  We also note that
each time delay distance in a sample is largely independent, such that
the scatter between measurements provides a self-contained test of
unknown systematics. \citet{Suy++13} show that unknown unknowns are
currently negligible with respect to known unknowns. At present, the
power of the method is limited by the small sample size of known
lensed quasars: there are only a few known lensed quasars suitable for
this experiment.

%%%%%%%%%%%%%%%%%%%%%%%%%%%%%%%%%%%%%%%

\begin{figure*}
  \renewcommand{\baselinestretch}{0.7} \centering
  \includegraphics[width=0.8\textwidth]{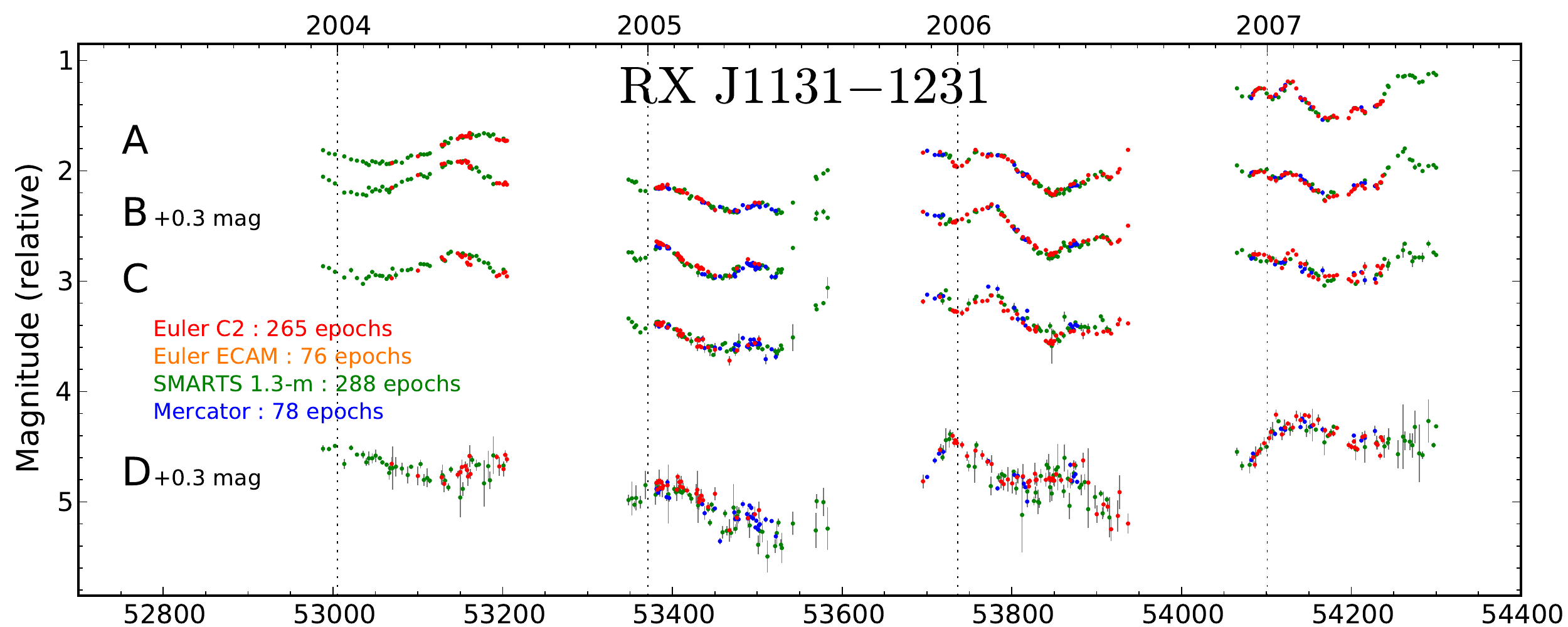}
  \includegraphics[width=0.8\textwidth]{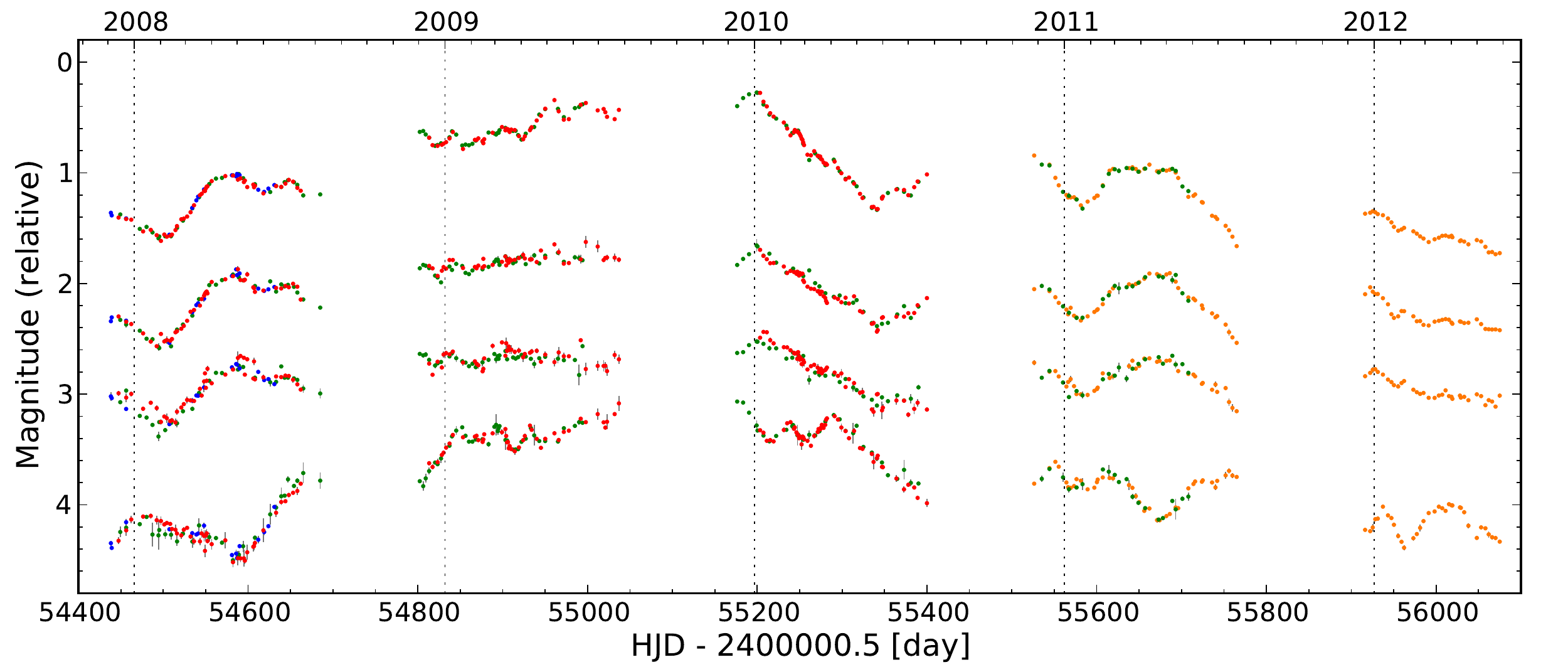}
  \caption{\label{fig:lightcurve}Example of multi-year monitoring of
  \rxj, yielding time delays with 1.5\% accuracy. The delay between
  images A,B,C and D can be clearly seen by eye! Figure from
  \citet{Tew++12}.}
\end{figure*}

\begin{figure*}
  \renewcommand{\baselinestretch}{0.7} \centering
  \includegraphics[width=0.45\textwidth]{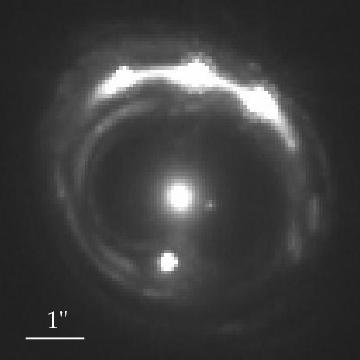}
  \includegraphics[width=0.45\textwidth]{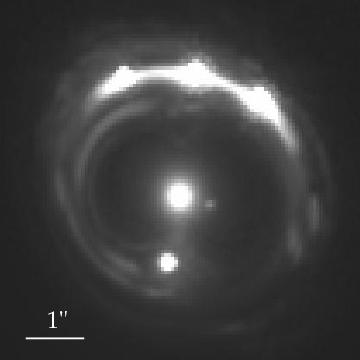}
  \caption{\label{fig:acsrecon} ACS image reconstruction of the lens
  system \rxj\ \cite{Suy++13}.  Left: observed ACS F814W image. Right:
  image predicted by the lens model. The lens model reproduces to high
  fidelity tens of thousands of data points providing extremely tight
  constraints on the mass model of the deflector and thus on
  cosmological parameters.}
\end{figure*}

\begin{figure*}
  \renewcommand{\baselinestretch}{0.7} \centering
  \includegraphics[width=0.4\textwidth,clip,
  angle=0]{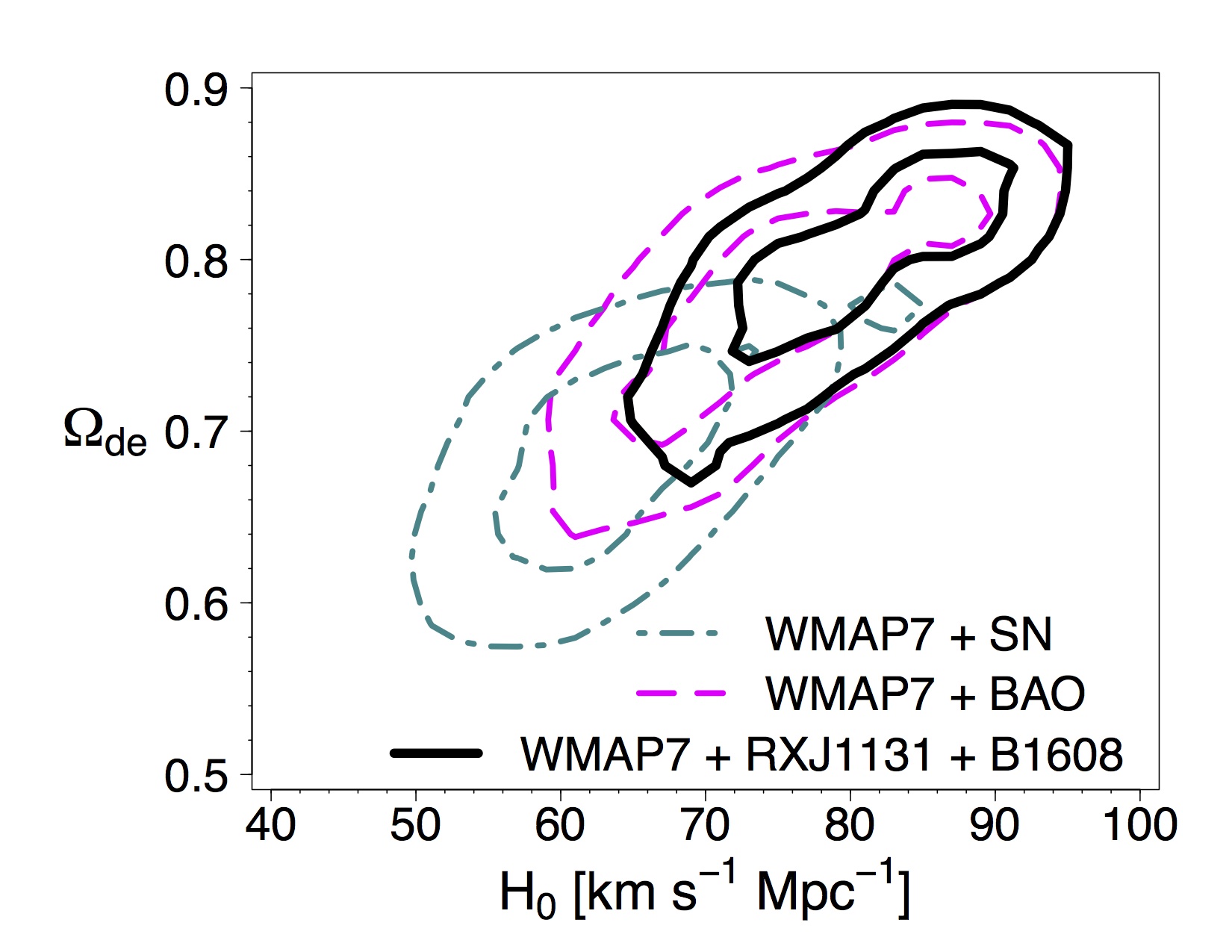}
  \includegraphics[width=0.4\textwidth,clip,
  angle=0]{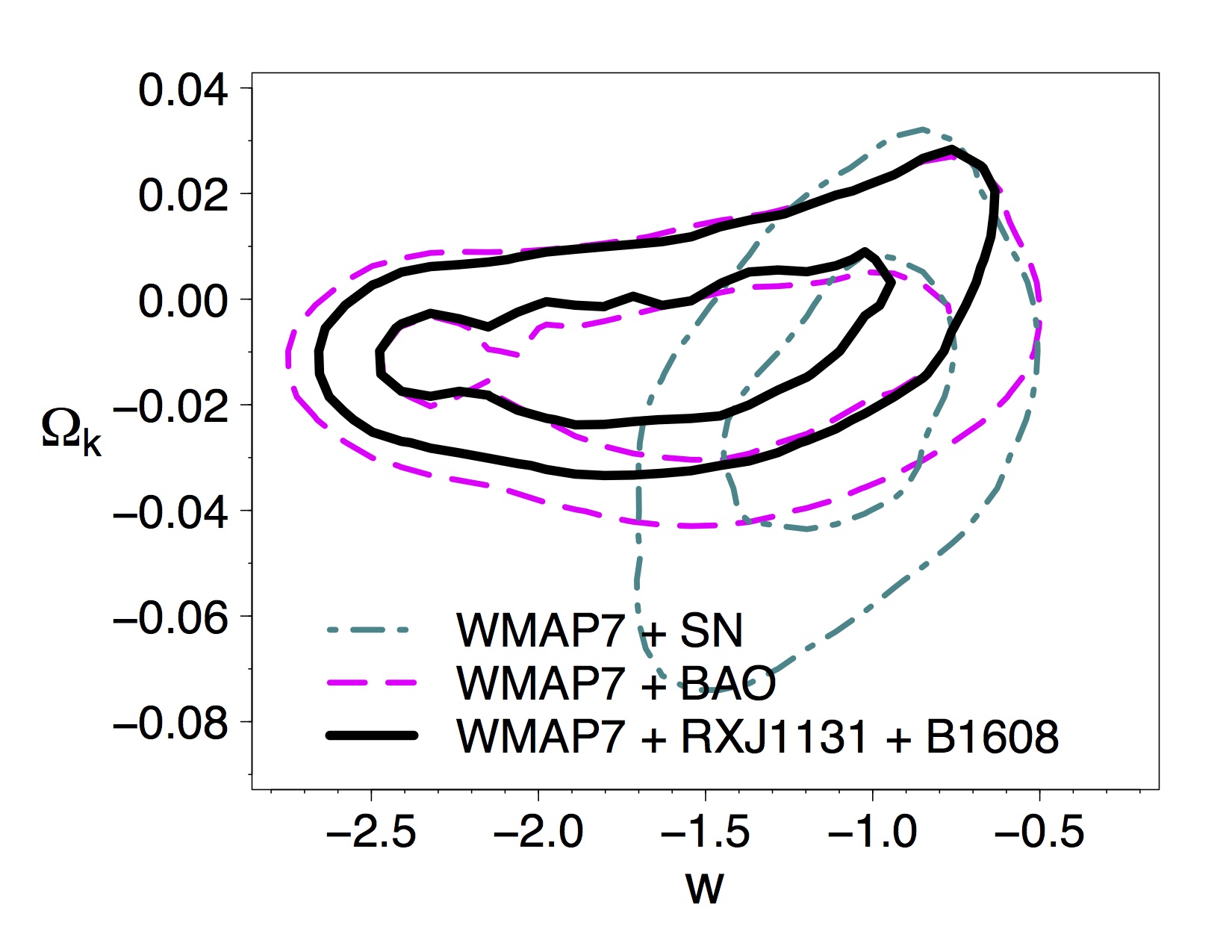}
  \caption{\label{fig:Lenses_BAO_SN} Posterior PDF of $H_0$, $\Ode$,
  $w$ and $\Ok$ for BAO (red dashed; \cite{PercivalEtal10}),
  Supernovae (SN) (blue dot-dashed; \cite{HickenEtal09}), time delay lenses
  (black solid; Suyu et al. 2013) when each is combined with WMAP7 in
  a curved cosmology with non trivial dark matter $ow$CDM.
  Contours/shaded regions mark the 68.3\%, and 95.4\% credible
  regions.  Time delay lenses are highly complementary to other
  probes, particularly CMB and SN. From \cite{Suy++13}.}
\end{figure*}

\section{Medium Term $(<2020)$ and Long Term (2020-2030) Goals}

A sample of $>100$ lensed quasars with well-measured time delays and
mass models will be transformative. As nicely summarized by
\citet{Lin11} ``Adding time delay data to supernovae plus cosmic
microwave background information can improve the dark energy figure of
merit by almost a factor of 5, and determine the matter density
$\Omega_m$ to 0.004, the Hubble constant $H_0$ to 0.7\% and the dark
energy equation of state time variation $w_a$ to $\pm0.26$,
systematics permitting.'' Furthermore, the analysis of this large
sample of systems will enable a direct check of unknown systematics by
comparing blindly the inferred parameters for each system. In this
white paper, we argue that this is achievable, through a concerted
observational effort, by the end of this decade.  Hundreds of lensed
quasars are expected to be discovered in ongoing Stage III imaging
surveys such as Pan-STARRS-1, the Dark Energy Survey (DES) and the
Subaru Hyper Suprime-Cam Survey
\citep{O+M10}. The human and observational resources required to confirm,
follow-up and derive cosmological parameters from this sample are
described below.

\begin{figure*}[h!]
  \renewcommand{\baselinestretch}{0.7}
  \centering
  \includegraphics[width=0.35\textwidth,clip]{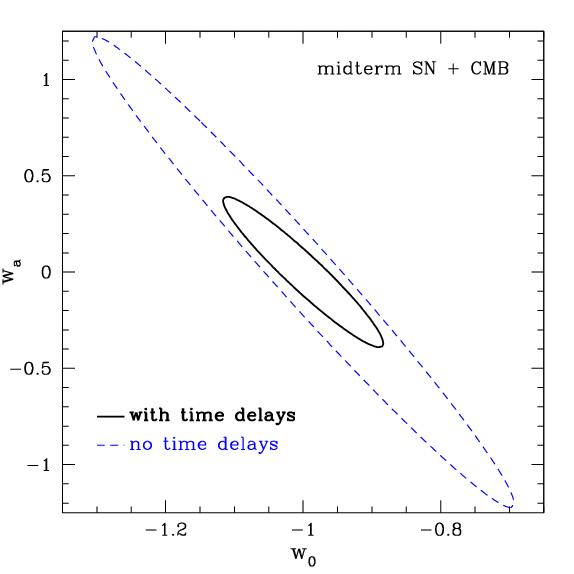}
  \includegraphics[width=0.35\textwidth,clip]{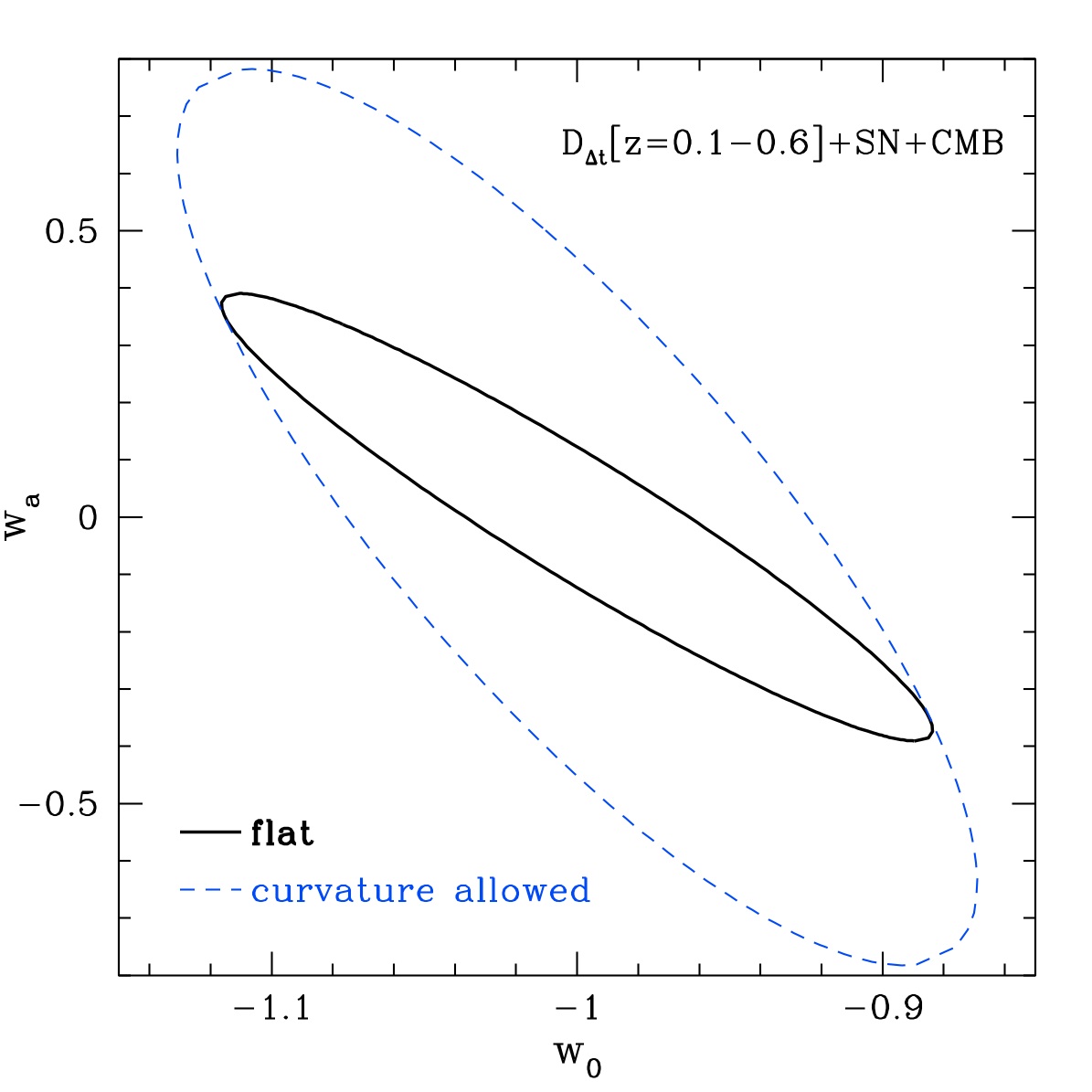}
  \caption{\label{fig:Linder} Forecasts for a time delay experiment
  from \cite{Lin11} based on 150 time delay distances. 
  %LSST should exceed this sample size by an order of magnitude. 
  %
  {\bf Left:}
  68\% confidence level constraints on the dark energy
  equation of state parameters $w_0$ and $w_a$ using midterm supernova
  distances and CMB information, and with (solid curve) or without
  (dashed curve) time delay measurements. The time delay probe
  demonstrates strong complementarity, tightening the area of
  uncertainty by a factor 4.8.
  {\bf Right}: 68\%
  confidence level constraints on the dark energy equation of state
  parameters $w_0$ and $w_a$ using time delay, midterm supernovae, and
  CMB information, assuming spatial flatness (solid curve) or allowing
  curvature (dashed curve). Although relaxing the flatness constraints
  worsens the constraints in the $(w_0,w_a)$ plane (there is a factor
  of 4 increase in the size of the confidence region between the solid
  and dashed curves), the constraints are much worse (a factor of 20
  larger than the solid curve) if the time delay data are not included.
  }
\end{figure*}

In the longer term, the LSST is expected to discover and provide
time delays for thousands of lensed quasars \citep{lsstwp}. This will
provide a further increase in sample size by over an order of
magnitude, necessary to continue to complement meaningfully  the
precision of other future cosmological experiments. As described in the
LSST Dark Energy Science Collaboration White Paper \citep{lsstwp}, a
number of challenges must be addressed in this decade in order for us
to be able to  take full advantage of the power of LSST. This is best
achieved through a combination of planned infrastructure work for LSST
and analysis of Stage III datasets.

\section{Requirements}

In order to measure time delay distances for each lensed quasar, the
following ingredients are needed: i) precise time delays; ii) deep
high resolution images of the lensed quasar host galaxy to model the
gravitational potential of the deflector; iii) the redshifts of both
the deflector and the source; iv) the stellar velocity dispersion of
the main deflector; v) multiband imaging of the field of the lens, and
redshifts of nearby companions, to characterize the environment along
the full line of sight.

The diversity of observational requirements characterizes time delay
cosmography. Whereas some other dark energy probes can be carried out
as self-contained experiment with a single facility (eg. BAO), or by
coupling a survey facility to a single follow-up facility (eg. SNeIa),
time delay lenses require a variety of existing facilities to
follow-up each planned survey.  We now discuss the facilities required for
each observational ingredient. The analysis requires dedicated
software, which is described in the last subsection.

\subsection{Time delays}

Recent work has demonstrated that multi-year monitoring campaigns are
needed to determine time delays with the necessary accuracy and
precision to carry out this experiment. A 5-year monitoring campaign
on a dedicated telescope is needed for each lens \citep{Tew++12}. Much
of the work is currently being carried out with 1m class telescopes
\citep[such as Euler and SMARTS;][]{Koc++06a,Tew++12}, but it has been difficult
to extend these campaigns, primarily because traditional observational
astronomy works on much shorter time scales. Time on general observer
telescopes is typically allocated on a semester by semester basis, and
then typically scheduled in contiguous blocks, preventing the
stability, longevity and flexibility necessary for this experiment.

With several 1-4m class telescopes currently being divested or closed,
there is an opportunity to transform some of these general observer
facilities into a dedicated dark energy experiment. What is required
is a robotic telescope (or network of) capable of delivering single
band images with typical seeing $\sim1'' $.  After an initial
investment for robotization (estimated at the level of a few
100k\$/telescope), the operating costs could be minimal given the
extremely simple program involved
($\sim$100k\$/yr/telescope). Depending on the size of the telescope,
only a fraction of the time might be sufficient.

After this decade, a dedicated monitoring system like this one could
continue to supplement the cadence provided by the LSST, while planned
space-based probes could provide higher-precision time delay
measurements in a smaller sample, sufficient to probe the nature of
dark matter as described in a companion white paper \citep[see
also][]{Mou++08}.

\subsection{High resolution imaging}

The necessary high-resolution imaging, with a stable and
well-characterized point-spread function, has so far been carried out
with the Hubble Space Telescope (HST). We expect HST to continue to be
a major workhorse for this application throughout the duration of its
mission. However, as the sample grows in size, obtaining high
resolution images would become a major bottleneck unless additional
capabilities are developed. After launch, the James Webb Space
Telescope (JWST) will provide some of this capability; in addition,
planned advances in adaptive optics technology will enable large
ground-based telescopes to complement JWST at bluer wavelengths.
High-strehl adaptive optics systems for 8-10m telescopes, and for the
next generation of extremely large telescopes, will provide higher
resolution than JWST and will be needed to fully exploit this method
into the next decade. For example, the Next Generation Adaptive Optics
system proposed for the Keck Telescope \citep{Max++08} and the NFIRAOS
system planned for the Thirty Meter Telescope \citep{NFIRAOS} will
provide this capability. With these advanced adaptive optics systems
it will be feasible to obtain high resolution imaging in a short amount
of time. For example, we estimate that complete imaging follow-up of
1000 lensed quasars discovered by LSST would take approximately 50
nights of Keck-NGAO, and 10 nights of Thirty Meter Telescope time.

\subsection{Redshifts of Lens and Source, and Lens Stellar Velocity
Dispersions}

Ground based spectroscopy is needed to determine redshifts of the
deflector and the source, as well as to measure the stellar velocity
dispersion of the main deflector (an additional independent mass
measurement important in constraining the normalisation of the density
profile). This is currently achieved using modest amounts of time on
8-10m class ground based telescopes \citep{T+K02b,Suy++13}. With the
continued operation of these telescopes through the next decade we
expect this to remain viable, although coordination with large scale
spectroscopic surveys, such as those planned for BAO experiments, may
be able to provide much of this information. AO systems such as those
mentioned above with integral field spectrographs could provide in one
shot a clean image of the source emission as well as the necessary
redshifts and velocity dispersions.

\subsection{Multiband Imaging of the Field, and Redshifts of Nearby
Companions, to Characterize the Lens Environment}

At present, determining the contribution of the environment and the
line of sight to the total convergence \citep{Tre10} requires
dedicated imaging and spectroscopic follow-up
\citep{Col++13,Gre++13}. As samples grow in size the expectation is
that this information will be provided by the finder surveys
themselves (e.g. DES in this decade, and LSST in the next).

\subsection{Software and analysis tools}

Time delay cosmography is computationally intensive at the moment,
typically requiring months of CPU {\it and} scientist-time for the
full exploration of the model uncertainties of each
system. Furthermore, characterization of the line of sight
contributions require ray-tracing through state of the art
cosmological simulations.  Both software requirements will be stressed
by the explosion of data expected in the near future. Exploiting this
dark energy probe will require advances in lens modeling software and
hardware to reduce the computing and investigator time per system to a
manageable level.  Likewise ray tracing through multiple independent
cosmological simulations with different cosmological parameters and
treatment of baryonic physics will be needed to quantify residual
systematic uncertainties below the 1\% level.

\section{Conclusions}

Recent studies have shown that gravitational time delays are a viable
tool for cosmography. The power of this method is currently limited by
the small size of current samples of known lensed quasars. However,
large samples will become available in the coming decade thanks to
dedicated surveys such as DES and LSST. The successful application of
this tool to incoming Stage III and Stage IV datasets thus hinges on
the development of follow-up capabilities. In turn, this requires the
continued support of a number of existing and approved facilities like
the James Webb Space Telescope and large ground based telescopes. In
addition to supporting these activities, we advocate the following
three priorities for new activities in the next two decades.

\begin{itemize} 
\item Support the development of software for the analysis of time delay systems and its application to existing and stage III datasets.
\item In this decade, we propose to convert a 2-4m class telescope
(or a network of 1m telescopes) into a dark energy monitoring
experiment. This would be transformative and cost effective when
compared with dedicated surveys or space missions, requiring funding to
cover only a fraction of the operating cost of a ground based imaging
telescope.
\item In the 2020 decade, the top priority will be to support the
development of diffraction-limited high performance adaptive optics
systems and their instrumentation on 8-30m class ground based
telescopes to enable high fidelity lens mass modeling.
\end{itemize}

\end{document}